\documentclass[twocolumn,floatfix,superscriptaddress,pra,aps,10pt,nofootinbib,longbibliography]{revtex4-2}
\usepackage{times}
\usepackage{amssymb}
\usepackage{color}
\usepackage{amsmath}
\usepackage{amsbsy}
\usepackage{amsthm}
\usepackage{graphicx}
\usepackage{bbm}
\usepackage{bm}
\usepackage{epsfig}
\usepackage{xfrac}
\usepackage{xcolor}
\usepackage{enumerate}
\usepackage{multirow}

\usepackage[normalem]{ulem}

\definecolor{myurlcolor}{rgb}{0,0,0.7}
\definecolor{myrefcolor}{rgb}{0.8,0,0}
\usepackage[unicode=true,pdfusetitle, bookmarks=true,bookmarksnumbered=false,bookmarksopen=false, breaklinks=false,pdfborder={0 0 0},backref=false,colorlinks=true, linkcolor=myrefcolor,citecolor=myurlcolor,urlcolor=myurlcolor]{hyperref}

\newcommand{\ket}[1]{\left| {#1} \right\rangle}
\newcommand{\bra}[1]{\left\langle {#1}\right|}

\newcommand{\braket}[2]{\langle #1|#2\rangle}

\renewcommand{\t}[1]{\textrm{#1}}
\newcommand{\tr}[0]{\mathrm{Tr}}

\newcommand{\real}{\mathrm{Re}}

\newcommand{\vspan}{\mathrm{span}}

\usepackage{braket}
\newcommand{\thmref}[1]{\hyperref[#1]{Theorem~\ref{#1}}}
\newcommand{\lemmaref}[1]{\hyperref[#1]{Lemma~\ref{#1}}}
\newcommand{\figref}[1]{\hyperref[#1]{Fig.~\ref{#1}}}
\newcommand{\figaref}[1]{\hyperref[#1]{Fig.~\ref{#1}a}}
\newcommand{\figbref}[1]{\hyperref[#1]{Fig.~\ref{#1}b}}
\newcommand{\figcref}[1]{\hyperref[#1]{Fig.~\ref{#1}c}}
\renewcommand{\eqref}[1]{\hyperref[#1]{Eq.~(\ref{#1})}}
\newcommand{\eqsref}[2]{\hyperref[#1]{Eqs.~(\ref{#1})-(\ref{#2})}}
\newcommand{\appref}[1]{\hyperref[#1]{Appx.~\ref{#1}}}

\newcommand{\aaa}{\mathfrak{a}}
\newcommand{\bbb}{\mathfrak{b}}

\newcommand{\var}{\varphi}

\newcommand{\F}{I}

\usepackage{dsfont}

\newtheorem{theorem}{Theorem}
\newtheorem{lemma}[theorem]{Lemma}

\newcommand{\wg}[1]{{\color{blue} #1}}

\usepackage{tabularray}

\begin{document}
\title{Interplay between time and energy in bosonic noisy quantum metrology}
\author{Wojciech G{\'o}recki}
\affiliation{INFN Sez. Pavia, via Bassi 6, I-27100 Pavia, Italy}
\author{Francesco Albarelli}
\affiliation{Scuola Normale Superiore, I-56126 Pisa, Italy}
\author{Simone Felicetti}
\affiliation{Institute for Complex Systems, National Research Council (ISC-CNR), Via dei Taurini 19, 00185 Rome, Italy}
\affiliation{Physics Department, Sapienza University, P.le A. Moro 2, 00185 Rome, Italy}
\author{Roberto Di Candia}
\affiliation{Department of Information and Communications Engineering, Aalto University, Espoo, 02150 Finland}
\affiliation{Dip. Fisica, University of Pavia, via Bassi 6, I-27100 Pavia, Italy}

\author{Lorenzo Maccone}
\affiliation{INFN Sez. Pavia, via Bassi 6, I-27100 Pavia, Italy}
\affiliation{Dip. Fisica, University of Pavia, via Bassi 6, I-27100 Pavia, Italy}

\begin{abstract}
Quantum entanglement and coherence often allow for protocols that outperform classical ones in estimating a system’s parameter.
When using infinite-dimensional probes (such as a bosonic mode), one could in principle obtain infinite precision in a finite time for both classical and quantum protocols, which makes it hard to quantify potential quantum advantage. However, such a situation is unphysical, as it would require infinite resources, so one needs to impose some additional constraint: typically the average energy employed by the probe is finite.
Here we treat both energy and time as a resource, showing that, in the presence of noise, there is a nontrivial interplay between the average energy and the time devoted to the estimation.
Our results are valid for the most general metrological schemes (e.g.~adaptive schemes which may involve entanglement with external ancillae or any kind of continuous measurement). We apply recently derived precision bounds for all parameters characterizing the paradigmatic case of a bosonic mode, subject to Lindbladian noise. We show how the time employed in the estimation should be partitioned in order to achieve the best possible precision. In most cases, the optimal performance may be obtained without the necessity of adaptivity or entanglement with ancilla.
We compare results with classical strategies. Interestingly, for temperature estimation, applying a fast-prepare-and-measure protocol with Fock states provides better scaling with the number of photons than any classical strategy.
\end{abstract}

\maketitle

\section{Introduction}

The use of nonclassical properties in quantum metrology, such as entanglement and squeezing, makes it possible to achieve a quadratic scaling of measurement precision with the amount of resources, known as Heisenberg scaling~\cite{giovannetti2006quantum,Paris2009,giovannetti2011advances,Toth2014,demkowicz2015optical,Schnabel2016,degen2017quantum,Pezze2018,Pirandola2018}.
For adequately high resource levels, it is possible to go beyond the shot noise limit.
Typically, in the presence of noise, Heisenberg scaling does not hold indefinitely as the amount of resources increases, and the quantum advantage comes down to a constant, whose exact value depends on the type and intensity of the noise~\cite{fujiwara2008fibre,escher2011general,demkowicz2012elusive,knysh2014true,demkowicz2014using,zhou2021asymptotic,kurdzialek2022using}.

\begin{figure}[t!]
\def\arraystretch{2.5}
\begin{tabular}{|c|c|c|c|}
\hline
Parameter             & Coherent light                            & Quantum bound                                  & Sec.                                 \\ \hline\hline
Frequency $\omega$    & $0.37\times\dfrac{4N}{\Gamma(1+2n_E)}$ & $\dfrac{4N}{\Gamma(1+2n_E)}$               & \ref{sec:freq}      \\ \hline
Displacement $\alpha$ & $0.82\times \dfrac{4}{\Gamma}$         & $\dfrac{4}{\Gamma}$                        & \ref{sec:disp}      \\ \hline
Squeezing $\epsilon$            & $-$                                   & $\dfrac{4(2N+1)}{\Gamma\sqrt{n_E(1+n_E)}}$ & \ref{sec:squeezing} \\ \hline
Loss rate $\Gamma$        & $0.37\times\dfrac{N}{\Gamma(1+2n_E)}$  & $\dfrac{N(1+2n_E)}{\Gamma}$                & \ref{sec:loss}      \\ \hline
Temperature $n_E$          & $\dfrac{\Gamma}{n_E}$                  & $N\times\dfrac{\Gamma(1+2n_E)}{n_E(1+n_E)}$      & \ref{sec:temp}      \\ \hline
\end{tabular}
\caption{Summary of the results. Optimal signal-to-noise ratio per time $S(t)/t$ for estimation of five different parameters.
We consider a single-mode bosonic system coupled to a thermal bath. $N$ is the mean number of photons, $\Gamma$ is the constant of coupling with the thermal bath (loss rate) and $n_E$ is the mean number of thermal photons. For simplicity of formulas, limits of large $N$ are taken.
In the column ``coherent light'' we present the results obtainable by classical strategy. In the column ``quantum bound'' we present a fundamental bound, that cannot be beaten by any quantum strategy (even including adaptiveness); in the main text, we present exemplary strategies saturating these bounds.
One can see that for temperature estimation quantum strategies give an advantage that scales with the number of photons, while for the other parameters the advantage is limited to a constant factor of order one ($0.37$ or $0.82$).
For squeezing estimation, there is no ``classical'' counterpart, as the Hamiltonian itself creates squeezing.}
\label{tab:summary}
\end{figure}

Often, one does not explicitly consider time as a resource in the estimation, but this is a crucial aspect when analyzing practical scenarios: it is quite obvious that, in general, by devoting more time to measurement, one can obtain an increase in precision, either by increasing the sensing time in each run or by repeating the experiment more times~\cite{helstrom1976quantum,giovannetti2004quantum}.
In most papers about fundamental metrological bounds including time, only small systems of non-scalable size are discussed~\cite{zhou2018achieving,das2024universal,das2024universal}, while the dependence of both time and energy/size appears only in~\cite{demkowicz2017adaptive,wan2022bounds}, where the general framework is provided.
Here we consider explicit case studies to show how time must be considered as a resource, emphasizing the comparison between classical and quantum strategies.
In general, it is not a priori clear how the available time should be allocated to get the optimal possible precision.
For example, consider the simple noiseless situation where one wants to estimate a parameter $\varphi$ using a unitary transformation $\exp(-it\var G)$ that acts on some probe state.
Then, the quantum Cramér-Rao bound (when applicable) gives $\Delta^2\varphi\geqslant 1/(4t^2\Delta^2 G)\gtrsim 1/t^2N^2$, in terms of the average ``energy'' $N$ and the time $t$, where $\Delta^2\varphi$ is the variance of any (unbiased) estimator, $\Delta^2G$ is the variance of the generator $G$ and we consider situations where $\Delta G\lesssim N$ (the opposite regime is useless for estimation \cite{giovannetti2012sub}).
For generic noise models, Heisenberg scaling is typically lost for both time and energy, leading to standard scaling $\sim const./tN$~\cite{demkowicz2017adaptive,wan2022bounds}, with a constant factor limiting any potential quantum advantage.
Optimization of the protocol requires consideration of the fact that often highly entangled states are more sensitive to noise, so one may take advantage of them only in a very short evolution.
As a result, a compromise must be found between taking more advantage from entanglement or from a longer time of evolution.

In this work we provide a systematic analysis of parameter estimation for a paradigmatic, experimentally relevant model: a single bosonic mode connected to a thermal bath.
We consider both Hamiltonian estimation (frequency, displacement, squeezing) and noise-parameter estimation (loss rate, temperature), all in the finite temperature regime.
We derive the fundamental bounds for these cases using the results in~\cite{demkowicz2017adaptive,wan2022bounds}, valid even for arbitrary strategies.
These include the use of ancillary degrees of freedom and adaptivity, e.g. continuous measurement and feedback~\cite{Nurdin2022,Rossi2020,Amoros-Binefa2021,Boeyens2023}, and exploiting quantum criticality~\cite{Garbe2020,cai2021observation,garbe2022critical,Gietka2022,DiCandia2023, alushi2024optimality,Cabot24,alushi2024collective}, as well as combining both approaches~\cite{Ilias2022,Salvia2023a,Cabot2024}.
Furthermore, we propose practical protocols that saturate the optimal bounds.
For most parameters, these optimal protocols only require feasible operations, readily available in relevant quantum platforms such as superconducting circuits, trapped ions, opto- and electro-mechanical systems, among others.

We show that for frequency and displacement estimation, almost optimal performance may be already obtained by using classical states of light if the total time is sufficiently large.
In particular, for frequency estimation we show that both a prepare-and-measure protocol and continuous measurement of the cavity's output field are optimal for long times.
Nonclassical light may still be useful in the case of a limited probing time.
On the contrary, for noise estimation a nonclassical input state grants a better scaling of the precision with temperature (for loss rate estimation) or with the number of photons in the cavity (for temperature estimation); here optimal performance may be obtained by a fast-prepare-and-measure protocol.
For squeezing estimation, there is no classical counterpart.
Importantly, in the limit of temperature going to $0$, Heisenberg scaling is not ruled out by the bounds.
In this limit, we propose a protocol based on error correction that achieves a variance $\sim const./ t N^2 $.
The results are summarized in \autoref{tab:summary}.


\section{General framework}


In this section we introduce the general framework and tools for noisy parameter estimation~\cite{demkowicz2017adaptive,wan2022bounds}.
While in the main part of the paper we will apply them to the single-mode bosonic model, here we keep the presentation more general and abstract, to stress the fact, that they can be directly used also to analyze other models.

\subsection{Problem formulation}
Consider a general  Lindblad evolution:
\begin{equation}
\label{eq:lind}
\frac{d\rho}{dt}=-i[H,\rho]+\sum_{j=1}^JL_j\rho L_j^\dagger -\frac{1}{2}\rho L_j^\dagger L_j-\frac{1}{2}L_j^\dagger L_j\rho,
\end{equation}
where $H$ and $L_j$ are the operators acting on the system
$\mathcal{H}_S$ which may depend on the unknown parameter $\var$ to
estimate.  Let us consider the most general metrological
protocol. We assume that the system may be entangled with a noiseless ancilla  $\mathcal H_A$ of arbitrary dimension (so that the  Hamiltonian and
Lindblad operators act on the joint Hilbert space as
$H\otimes\openone_A$ and $L_j\otimes\openone_A$ respectively);
moreover, we allow for an additional unitary $V(t)$ (acting jointly on
$\mathcal H_S\otimes\mathcal H_A$) which is fully flexible and
controllable. Finally, the observable $\hat O$ is
measured on the joint output state.
Note that such general and abstract formulation covers also situations where a weak measurement of the system is performed continuously in time and future action on the system depends on previous measurement outcomes (all these procedures can be mathematically simulated as joint unitary transformations of the system with ancilla, followed by the measurement at the end).

We consider the quantum Fisher information (QFI) of the output state, which is equal to the signal-to-noise ratio (SNR) optimized over the possible choices of the measured observable:
\begin{equation}
\label{eq:fishdef}
\F_\var=\max_{\hat O}S,\quad \t{where}\quad S=\frac{|\partial_\var \braket{\hat O}|^2}{\Delta^2\hat O}.
\end{equation}
(From the point of view of the local estimation, optimization over a single observable is equivalent to joint optimization over measurement and estimator, see App. \ref{app:QFI} for a formal derivation and \autoref{sec:noise} for further discussion).
We assume an ideal scenario, where both state preparation and the measurement procedure may be performed instantaneously, so only sensing takes time.
Therefore, any bound derived for such an ideal scenario will still be valid for any more realistic situation.


If a quadratic scaling in time can be sustained indefinitely, the optimal strategy is to use all available time coherently in a single realization in the experiment.
However, in the occurrence of noise, for typical protocols, the QFI grows only up to a certain time.
Therefore, after that, it is better to interrupt the measurement and re-start the procedure from the beginning.
The QFI scales linearly with the number of repetitions.
Having at our disposal a total time $T$ and spending time $t$ for a single iteration, in principle, we are able to perform $T/t$ repetitions, so the total QFI will be equal to $(T/t)\cdot \F_\var(t)$.
Therefore, to validate the usefulness of a given protocol (properly taking into account time as a resource) one should compare $\max_t \F_\var(t)/t$.
This figure of merit is routinely employed in noisy quantum metrology~\cite{wan2022bounds}, such as frequency estimation~\cite{huelga1997improvement,chin2012quantum,delcampo2013quantum,Haase2017} and thermometry~\cite{correa2015individual,Sekatski2022optimal,Albarelli2023,Mirkhalaf2024}.

\subsection{Noisy bounds}
Let us start with an estimation problem where the parameter dependence is in the Hamiltonian, and denote the derivative with respect to this parameter as $\dot H:=\partial_\var H$.
In the absence of noise, the QFI is bounded by~\cite{garbe2022critical}:
\begin{equation}
\label{eq:quadratic}
\F_\var(t)\leq 4\left[\int_0^t \sqrt{\Delta^2 \dot H}dt'\right]^2.
\end{equation}
In particular, for the simple case $H=\var G$, we recover the standard formula $I_\var(t)\leq 4t^2\Delta^2G$.

While this bound is derived for noiseless Hamiltonian dynamics, it applies also in the presence of noise.
In this case,
it can typically be saturated for sufficiently small times, when the evolution is almost unitary.

However, for most generic noise, quadratic scaling with time occurs only at the beginning, while for longer times the scaling of the QFI becomes linear.
More formally, quadratic scaling for large times is possible if and only if the generator of the evolution cannot be expressed as a quadratic combination of Lindblad operators
\begin{equation}
\label{eq:hnls}
 \dot{H} \notin \t{span}_{\mathbb C}\{\openone, L_i, L^\dagger_i, L^\dagger_iL_j\}.
\end{equation}
This is the so-called Hamiltonian-not-in-Lindblad-Space (HNLS) condition \cite{zhou2018achieving}.
Otherwise, the QFI is bounded by \cite{demkowicz2017adaptive,wan2022bounds}:
\begin{equation}
\label{eq:linear}
\F_\var(t)\leq 4\int_0^t \braket{\aaa(h)}_{t'}dt',\quad 
\t{with}\quad \bbb(h)=0,
\end{equation}
where $\braket{\cdot}_{t'}:=\tr(\rho(t')\cdot)$, and  $\aaa(h)$ and $\bbb(h)$ are the following operators acting on the system:
\begin{eqnarray}
\label{eq:aoperator}
\mathfrak{a}(h) & = & {\left(\mathfrak{h} \vec{L} + \vec{h} \openone \right)}^\dagger\cdot \left(\mathfrak{h} \vec{L} + \vec{h} \openone\right),\\
\label{eq:boperator}
\mathfrak{b}(h) & = &\dot{H} + h_{00}\openone+\vec{h}^\dag\cdot\vec{L}+\vec{L}^\dag\cdot\vec{h}+\vec{L}^\dag \cdot\mathfrak{h}\cdot\vec{L}.
\end{eqnarray}
Here $\vec{L}$ is a vector of Lindblad operators, while the letter $h$ jointly denotes all the free parameters: a real scalar $h_{00} \in \mathbb{R}$, a complex vector $\vec{h} \in \mathbb{C}^J$, and a hermitian matrix $\mathfrak{h} \in \mathbb{C}^{J \times J}_{\t{H}}$.
The bound is valid for any $h$ satisfying $\bbb(h)=0$ and may be tightened by optimization over $h$.

The time dependence of $\Delta^2 \dot H$ and $\braket{\aaa(h)}_{t'}$ may be essential in analyzing critical metrology, where the average values of observables and their variances are changing significantly during evolution. For the passive strategies (meaning that no additional action is performed during the evolution, and a measurement is performed at the end), they do not change their order of magnitude and comparing the bounds \eqref{eq:quadratic} and \eqref{eq:linear} gives the characteristic time for which the scaling goes from quadratic to linear $\tau\sim{\braket{\aaa(h)}}/{\Delta^2 \dot H}$~\cite{das2024universal}.
In the intermediate regime, slightly tighter bounds may be derived~\cite{wan2022bounds}, see App.~\ref{app:general} for details.

If the total available time is much longer than $\tau$, then, according to the discussion from the previous section, in validating the measurement protocol it is reasonable to look at the ratio $\F_{\var}(t)/t$. For simplicity of further formulas, we may also bound the integral in \eqref{eq:linear} by the maximal value of the function under the integral to obtain:
\begin{equation}
\label{eq:main}
\frac{\F_\var(t)}{t}\leq 4\braket{\aaa(h)}_{\max},\quad\t{with}\quad \bbb(h)=0,
\end{equation}
where $\braket{\aaa(h)}_{\max}:=\max_{t'\in[0,t]}\braket{\aaa(h)}_{t'}$. This inequality will be the main point of reference in discussing examples.

\section{Hamiltonian parameter estimation}

We analyze several specific estimation strategies for a single-mode system coupled to a thermal bath:
\begin{multline}
\label{eq:model}
\frac{d\rho}{dt}=-i[H,\rho]+
\Gamma (1+n_E)\left(a\rho a^\dag -\frac{1}{2}\{a^\dag a,\rho\}\right)\\+\Gamma n_E\left(a^\dag\rho a -\frac{1}{2}\{a a^\dag,\rho\}\right),
\end{multline}
so the Lindblad operators are $L_1=\sqrt{\Gamma(1+n_E)}a$ and 
$L_2=\sqrt{\Gamma n_E}a^\dagger$.

\subsection{Frequency estimation}
\label{sec:freq}

We start with the problem of frequency estimation, i.e. the Hamiltonian is given as:
\begin{equation}
H=\omega a^\dagger a,
\end{equation}
and $\omega$ is the parameter to be estimated. The bound \eqref{eq:main} reads (see App. \ref{app:calculations} for derivation):
\begin{equation}
\label{eq:qfifreq}
\frac{\F_{\omega}(t)}{t}\leq\frac{4\braket{a^\dag a}_{\max}}{\Gamma(1+2n_E-\frac{n_E}{\braket{a^\dag a}_{\max}+1})}\approx\frac{4\braket{a^\dag a}_{\max}}{\Gamma(1+2n_E)},
\end{equation}
so it scales linearly with the number of photons.
Note that, for arbitrary states, the mean number of photons does not impose any bound on the variance, which appears in the quadratic bound~\eqref{eq:quadratic}, so without additional assumptions we cannot conclude anything about the characteristic transition time $\tau$.
However, restricting to Gaussian states we have $\Delta^2 N\leq 2N(N+1)$, and then:
\begin{equation}
    \tau \gtrsim\frac{1}{2(N+1)\Gamma (1+2n_E)}.
\end{equation}
To understand it better, consider an input state to be a vacuum squeezed along the $x$ axis, shifted along the $p$ axis $\mathcal D(i\alpha)S(r)\ket{0}$, where $\mathcal D$ and $S$ are the displacement and squeezing operators,  so the initial number of photons is $N=\alpha^2+\sinh^2 r$, with $\alpha \in \mathbb{R}$.
Without loss of generality, we may assume that we work around the point $\omega\approx 0$ (otherwise we rotate the frame of reference).
After rotating for a time $t$, one performs homodyne detection of the observable $\hat x=a+a^\dagger$, which leads to:
\begin{equation}
\begin{split}
\braket{\hat x(t)}&=2\alpha e^{-\Gamma t/2}\sin(\omega t)\\
\Delta^2 \hat x(t)&=e^{-\Gamma t}e^{-2r}+(1-e^{-\Gamma t})(1+2n_E),
\end{split}
\end{equation}
and the SNR is equal to
\begin{equation}
\label{eq:snrfeq}
S_\omega(t)=\frac{4\alpha^2t^2}{e^{-2r}+(e^{\Gamma t}-1)(1+2n_E)}.
\end{equation}
This formula gives a very simple interpretation of the scaling with both $t$ and $N$, which have different origins.
For short times (compared to the loss rate), the signal is coherently accumulated over time, resulting in $t^2$ scaling of the SNR.
In the case of photons, the light intensity $\alpha^2$ multiplies the signal strength, while the squeezing factor $e^{-2r}$ is responsible for reducing the noise.
In the unitary case $\Gamma=0$, it is enough to evolve state freely to obtain a quadratic scaling with time, while obtaining a quadratic scaling with $N$ requires an equal distribution of achievable photons between $\alpha^2$ and $\sinh^2r$.

\begin{figure}[t!]
  \includegraphics[width=0.48
\textwidth]{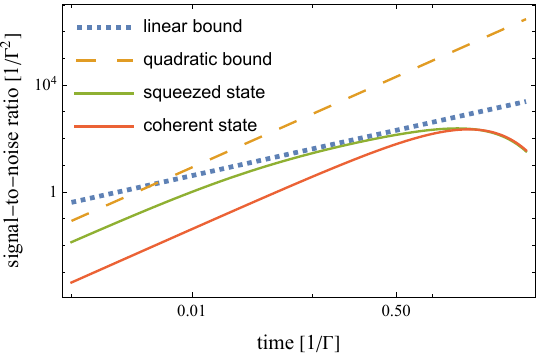}
\centering
\caption{Relation between exemplary metrological strategies and fundamental bounds for precision on the example of frequency estimation.
Both the quadratic one~\eqref{eq:quadratic} and the linear one~\eqref{eq:linear} are valid during the whole evolution, for any strategy of given average energy.
The quadratic one is tighter for $t\ll 1/\Gamma$ and the linear one is tighter for $t\gtrsim 1/\Gamma$.
Comparing specific strategies, we see that for small times the squeezed state performs significantly better than the coherent one.
However, it also transitions from quadratic to linear scaling faster.
At the optimal time, they perform similarly, both close to the bound.}
\label{fig:fre}
\end{figure}

For $\Gamma>0$, if the time $t$ is strictly limited, obtaining the optimal scaling with the system parameters requires using a nonclassical input state.
This may happen, for instance, if the signal appears only for some limited time, of order $1/\Gamma$ or smaller.
The optimal rate of squeezing then depends on the exact values of $t,N,\Gamma$. 
In the limit of large $N$, for $1/N\ll e^{-2r}\ll \Gamma t \ll 1$, approximately the whole energy is put into the displacement $N\approx \alpha^2$, and \eqref{eq:snrfeq} divided by time converge to \eqref{eq:qfifreq}.
Note, however, that if the total time $t\gg 1/\Gamma$ and one has the freedom to divide it into optimal slots, very similar results may be obtained by using a coherent state (so $r=0$) and letting it evolve for a time $1/\Gamma$ in each iteration.
The precision of this classical protocol loses at most a factor $1/e$ compared to the optimal bound.
Exactly the same factor has been observed for phase-shift estimation, when comparing sequential vs. parallel use of $N$ quantum gates~\cite{demkomaccone2014}.
For finite temperature $n_E>0$, the gap between using only coherent states and the optimal bound is even smaller.
Indeed, using a squeezed state significantly reduces the variance of measurement results, but this effect disappears sharply for times longer than $e^{-2r}/\Gamma$.
On the contrary, the optimal signal accumulation requires times of order $1/\Gamma$.
As a consequence, one must choose if one wants to take more benefits from entanglement between photons or from coherence in time, but one cannot use both of them simultaneously, see~\autoref{fig:fre}.
Note that here we assume that $n_E$ is a constant independent on $\omega$.
This is a common and good approximation, as typically the knowledge about $\omega$ which could be taken from measuring $n_E$ is negligible.
Yet, one should remember that, in reality, a more reasonable assumption is $n_E=1/(e^{\hbar\omega/k_BT}-1)$, which becomes relevant if the measurement is performed on a thermal state~\cite{ostermann2023temperature}.

\begin{figure}[t!]
  \includegraphics[width=0.48
\textwidth]{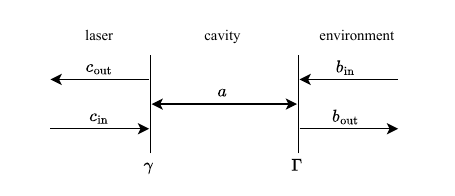}
\centering
\caption{A schematic representation of the cavity field and the input and output fields for a
double-sided cavity. Environment input mode $b_{\t{in}}$ is in the vacuum state (coupled with constant $\Gamma$), while laser input mode $c_{\t{in}}$ is in the coherent state (coupled with constant $\gamma$).
}
\label{fig:cav}
\end{figure}

In order to show that these results are relevant in a more general context, we now analyze the results for $t\gg 1/\Gamma$ in a less idealized scenario. 
We consider a protocol to estimate the frequency of a resonator by driving it with a coherent field and continuously measuring the reflected signal.
We model this situation as a two-sided cavity \cite[Sec. 7.3]{walls2008quantum} where, in addition to coupling to the modes $b_{\t{in/out}}$ of a dissipative environment with coupling constant $\Gamma$, the cavity is coupled to a set of input-output modes $c_{\t{in/out}}$ with coupling $\gamma$, see \figref{fig:cav}.
For simplicity, we assume that the environment is in a vacuum state.
We assume that the decay rate $\Gamma$ is fixed (and in practice already reduced as much as possible), and that no measurement can be implemented on the modes of the dissipative environment, which are thus responsible for the Lindbladian dissipator \eqref{eq:model}.
On the contrary, we assume that the coupling $\gamma$ to the input-output modes can be controlled and that arbitrary signals can be sent as input.
We stress that the bound \eqref{eq:qfifreq} remains valid in this physical context, since the input-output modes $c_{\t{in/out}}$ can be understood as the ancillary system already included in the derivation of the bound.
Importantly, the mean number of photons appearing in the bound $\braket{a^\dagger a}_{\t{max}}$ refers only to the photons inside the resonator, not to the total power of the input signal. 

We assume the cavity is driven with coherent monochromatic light with photon flux $|\alpha_f|^2$ and frequency $\omega_0$.
The mean number of photons in the cavity for the steady state is then (see App. \ref{app:cont}):
\begin{equation}
N_{ss}:=\braket{a^\dagger a}_{ss}=\frac{\gamma|\alpha_f|^2}{ \bigl(\frac{\Gamma+\gamma}{2}\bigr)^2+(\omega-\omega_0)^2}.
\end{equation}
We estimate the frequency from the number of photons in the output mode.
From energy conservation, the output photon flux in the mode $c_{\t{out}}$ is equal to $n_{\t{out}}=|\alpha_f|^2-\Gamma N_{\t{ss}}$.
During a total time $t$ we therefore observe $N_{\t{out}}^t=\int_0^t n_{\t{out}} \, dt'$ photons on average.
The signal-to-noise ratio takes its maximum at the point $\gamma\approx \Gamma$ and $\omega\approx \omega_0$ (more precisely, in the regime $\Gamma^2\gg (\omega-\omega_0)^2\gg (\Gamma-\gamma)^2$), and reads:
\begin{equation}
\label{eq:snrcont}
S_\omega(t)=\frac{|\partial_\omega N_{\t{out}}^t|^2}{\Delta^2 N_{\t{out}}^t}\approx\frac{4N_{ss}t}{\Gamma}.
\end{equation}
Here, the variable $t$ represents the sensing time, which starts \emph{after} having reached the steady state in the cavity.
In principle, the total duration of the protocol should also take into account the time needed to reach the steady state (as it requires interacting with the system as well).
Nevertheless, if the total time is far larger than the relaxation time of the system $t\gg 1/\Gamma$, the time spent to obtain the steady state is negligible.
Thus, the SNR \eqref{eq:snrcont} saturates the bound \eqref{eq:qfifreq} for large times,
showing that a coherent state input and time-continuous photodetection of the reflected signal are optimal when $t\gg 1/\Gamma$.
Otherwise, if the total time is constrained to be shorter than the relaxation time, even in this scenario involving control and readout of the system through input-output modes, the optimal strategy involves squeezing.

It should be stressed that for $\omega\approx\omega_0$ and $\gamma \approx \Gamma $, the expectation value of the output photons in mode $c_{\t{out}}$ is around zero, which makes the protocol very unstable for even minimal changes of the coupling constant or any additional uncertainty induced during the measurement~\cite{gorecki2022quantum,kurdzialek2023measurement,girotti2024optimal}. 
From a practical point of view, it would be more stable to work around the point $|\omega-\omega_0|\sim \Gamma$, for which the SNR still gives a linear scaling in $N_{ss}t/\Gamma$, but with a different proportionality constant smaller than $4$, see App. \ref{app:cont} for the full derivation and a more detailed discussion. 

The bound guarantees that the asymptotic linear scaling cannot be overcome even with more advanced methods, such as coherent or measurement-based feedback or a nonclassical (entangled or squeezed) input field.
For example, in~\cite{alushi2024optimality} the additional squeezing term $+\epsilon(a^{\dag 2}+a^2)$ has been added to the Hamiltonian in \eqref{eq:model}, which results in critical behavior around the point $\epsilon=\sqrt{\omega^2+\Gamma^2}$.
As shown, it allows for arbitrary lengthening of quadratic scaling time, while simultaneously reducing the scaling constant and effectively leading to the same scaling $\sim N t/\Gamma$ in the long-time limit.

\subsection{Displacement estimation}
\label{sec:disp}
We now turn to displacement estimation, i.e.
\begin{equation}
\label{eq:dispH}
H=i\alpha(a^\dag-a),
\end{equation}
where we assume, that the displacement direction is known, so that the parameter to be estimated is just the real coefficient $\alpha$.
The bound \eqref{eq:main} reads (see App. \ref{app:calculations}):
\begin{equation}
\label{eq:disp}
\frac{\F_\alpha(t)}{t}\leq \frac{4}{\Gamma (1+2n_E)}.
\end{equation}
Unlike the previous example, here the constraint on a mean number of photons implies a bound on $\Delta^2\dot H$, $\Delta^2\dot H\leq 4N+2$, so:
\begin{equation}
    \tau\gtrsim\frac{1}{(4N+2)\Gamma (1+2n_E)}.
\end{equation}

We see that the precision bound \eqref{eq:disp} does not scale with $N$ at all, differently from the previous case.
It may be easily understood, since in this model there is no point in using displacement on the input state
(as it commutes with the Hamiltonian), while squeezing, similarly to the previous case, may only lead to a constant advantage.
Indeed, for squeezed vacuum as input we obtain:
\begin{equation}
\begin{split}
\braket{\hat x(t)}&=2\alpha(1-e^{-t\Gamma/2})/(\Gamma/2),\\
\Delta^2 \hat x(t)&=e^{-\Gamma t}e^{-2r}+(1-e^{-\Gamma t})(1+2n_E),
\end{split}
\end{equation}
and the SNR per time is:
\begin{equation}
\frac{S_\alpha(t)}{t}=\frac{16(1-e^{t\Gamma/2})^2}{\Gamma^2(e^{-\Gamma t}e^{-2r}+(1-e^{-\Gamma t})(1+2n_E))t}.
\end{equation}
Again, the bound may be asymptotically saturated if $e^{-2r}\ll \Gamma t \ll 1$.
Similarly to frequency estimation, almost optimal performances may be also obtained without squeezing, by choosing the optimal time of a single iteration, leading finally to a factor $\sim3.26$ instead of $4$ in the nominator.

\subsection{Squeezing estimation}
\label{sec:squeezing}

In this section, we consider the estimation of the squeezing parameter (see also \cite{souza2024classes}).
Consider a system coupled to a thermal bath \eqref{eq:model}, with the Hamiltonian:
\begin{equation}
H=\epsilon (a^2+a^{\dagger 2}),
\end{equation}
where the parameter $\epsilon$ is to be measured. The bound \eqref{eq:main} reads:
\begin{equation}
\frac{\F_\epsilon(t)}{t}\leq \frac{4(2N+1)}{\Gamma \sqrt{n_E(1+n_E)}}.
\end{equation}
Note that the bound goes to infinity as $n_E\to 0$.
Indeed, if the bath is at zero temperature $n_E=0$,  we have only a single Lindblad operator $L=\sqrt{\Gamma}a$ and the condition $\mathfrak b(h)=0$ cannot be satisfied for any choice of $h$, so in principle, a quadratic scaling with both number of photons and total time $\F_\epsilon\sim N^2t^2$ should be possible, even for arbitrarily long times.

From the theoretical point of view such scaling indeed may be obtained by applying a proper quantum error correction protocol \cite{zhou2018achieving}, see App. \ref{app:squeezing} for details.
In practice, its implementation for this model would pose many difficulties, mainly due to the need for frequent projections onto non-trivial subspaces of Hilbert space.
Moreover, when dealing with Heisenberg scaling in both time and energy, another issue arises, related to the fact that saturating the Cramér-Rao bound demands arbitrarily large prior information~\cite{Giovannetti2012beyond,hall2012universality,Gorecki2020pi}.

Instead, we discuss here in detail a simpler protocol, leading to linear scaling with time, but providing quadratic scaling with the average number of photons in the system $\F_\epsilon\sim N^2t/\Gamma$ (in contrast to the examples discussed earlier, where such scaling was fundamentally forbidden).

The idea of this protocol is based on the observation that the evolution of the state coupled to a thermal bath may be seen as the deterministic decay followed by random quantum jump ~\cite{dalibard1992wave}.
This allows us to use a bosonic error correction code based on cat states 
~\cite{cochrane1999macroscopically,leghtas2013hardware}.
Below we follow the notation from~\cite{leghtas2013hardware}.
Consider cat states of the form:
\begin{equation}
\begin{split}
    \ket{\mathcal C_\alpha^\pm}&=\mathcal N(\ket{\alpha}\pm\ket{-\alpha}),\\
     \ket{\mathcal C_{i\alpha}^\pm}&=\mathcal N(\ket{i\alpha}\pm\ket{-i\alpha}).
    \end{split}
\end{equation}
For large $\alpha$, all the states $\ket{\alpha}$, $\ket{-\alpha}$, $\ket{i\alpha}$ and $\ket{-i\alpha}$ are almost orthogonal and the normalization constant becomes $\mathcal N\approx \frac{1}{\sqrt{2}}$.
A logical qubit may be encoded as $\ket{\psi_\alpha^{(0)}}=c_g\ket{\mathcal C_\alpha^+}+c_e\ket{\mathcal C_{i\alpha}^+}$.
Note that the action of subsequent quantum jumps results in the following states:
\begin{equation}
\begin{split}
   \ket{\psi_\alpha^{(0)}}:=&c_g\ket{\mathcal C_\alpha^+}+c_e\ket{\mathcal C_{i\alpha}^+},\\
   a\ket{\psi_\alpha^{(0)}}\propto  \ket{\psi_\alpha^{(1)}}:=&c_g\ket{\mathcal C_\alpha^-}+ic_e\ket{\mathcal C_{i\alpha}^-},\\
   a^2\ket{\psi_\alpha^{(0)}}\propto \ket{\psi_\alpha^{(2)}}:=&c_g\ket{\mathcal C_\alpha^+}-c_e\ket{\mathcal C_{i\alpha}^+},\\
   a^3\ket{\psi_\alpha^{(0)}}\propto \ket{\psi_\alpha^{(3)}}:=&c_g\ket{\mathcal C_\alpha^-}-ic_e\ket{\mathcal C_{i\alpha}^-}.
\end{split}
\end{equation}
Importantly, each of them has well-defined photon parity, which is changed after each quantum jump.
Therefore, by applying non-demolition parity measurement~\cite{leghtas2013hardware}, one may efficiently keep track of the actual states, without affecting the relative quantum phase.
Moreover, due to the deterministic decay, for all states $\alpha\to e^{-\Gamma t/2}\alpha$ in time.

Let us now look at how useful this protocol can be for squeezing estimation. 
Let us define a proper even (+) and odd (-) code spaces as $H_{\mathcal C^{\pm}}:=\vspan\{\ket{\mathcal C_\alpha^\pm},\ket{\mathcal C_{i\alpha}^\pm}\}$. The Hamiltonian acts non-trivially inside of both these code spaces:
\begin{multline}
\label{eq:HC}
H_{\mathcal C^{\pm}}:=
\Pi_{\mathcal C^{\pm}}
\epsilon (a^{\dag2}+a^2)\Pi_{\mathcal C^{\pm}}\\
=2\epsilon\real(\alpha^2)\left(\ket{\mathcal C_\alpha^{\pm}}\bra{\mathcal C_\alpha^{\pm}}-\ket{\mathcal C_{i\alpha}^{\pm}}\bra{\mathcal C_{i\alpha}^{\pm}}\right).
\end{multline}
Therefore, the evolution of the state $\ket{\psi_\alpha^{(0)}}=\frac{1}{\sqrt{2}}(\ket{\mathcal C_\alpha^+}+\ket{\mathcal C_{i\alpha}^+})$ will be sensitive to changes of $\epsilon$. As we noted, the noise does not take us outside of these two subspaces.

The only potential problem is that the acting of the measured Hamiltonian may create a leakage outside of them.
Nevertheless, since for large $|\alpha|^2$ the leading term $a^\dagger\ket{\alpha}$ may be approximated by $\alpha^*\ket{\alpha}$, the amount of leakage will remain negligible for $t \epsilon\ll 1/\sqrt{N+2}$, see App. \ref{app:catstate} for details.
Note especially, that this condition may be satisfied even for large $N$, if one has approximate knowledge about the value of the squeezing parameter.
As a matter of fact, one can apply anti-squeezing as well, to keep the effective squeezing parameter to be estimated around $0$.

Assuming that the above condition is satisfied (i.e. the effect of leakage may be neglected), the evolution of the states will be given as:
\begin{equation}
\ket{\psi(t)}=\frac{1}{\sqrt{2}}\left[
e^{-i\var(t)}\ket{\mathcal C_{\alpha e^{-t\Gamma/2}}^{(-1)^n}}+e^{i\var(t)}i^n\ket{\mathcal C_{i\alpha e^{-t\Gamma/2}}^{(-1)^n}}\right],
\end{equation}
where $n$ is the number of jumps and $\var(t)$ is the phase accumulated during evolution:
\begin{equation}
\var(t)=\int_0^t 2\epsilon\real(\alpha(t')^2)dt'=2\epsilon\alpha^2(1-e^{-t\Gamma})/\Gamma.
\end{equation}
As the number of jumps $n$ is monitored,
one effectively deals with a single qubit. Therefore $I_\epsilon(t)=16 N^2 (1-e^{-t\Gamma})^2/\Gamma^2$, so
\begin{equation}
\max_t\frac{I_\epsilon(t)}{t}=6.56\times\frac{N^2}{\Gamma},
\end{equation}
which is obtained for $t=1.26/\Gamma$.

\section{Noise estimation}
\label{sec:noise}
We now move beyond Hamiltonian parameters, and focus on noise estimation
i.e., we assume that the parameter is encoded solely in the dissipative part of the Lindblad equation~\eqref{eq:model}.
Moreover, we assume, that the parameter does not change the ``type'' of noise, but only its magnitude, i.e., the parameter is encoded in the scalars multiplying Lindblad operators.
Under this assumption, the QFI grows at most linearly with time and it is bounded by~\cite{wan2022bounds}:
\begin{equation}
\label{eq:noisebound}
\frac{\F_\var(t)}{t}\leq\braket{\dot{\vec{L}}^\dagger\dot{\vec{L}}}_{\max}.
\end{equation}
This setting corresponds to estimation of the temperature of thermal bath and of the loss rate.
For more general cases, where the parameter affects the shape of Lindblad operators, see App.~\ref{app:general}.
In contrast to unitary evolution, for which the short-time QFI grows quadratically, and for which a specific optimal time of a single iteration is required for optimal performance, for noise estimation the optimal time may be arbitrarily short.
We only need to remember that the Lindblad equation itself comes from averaging over some finite time so that the analysis for the smaller time scales is trivially nonmeaningful (as it would require a microscopic description that goes beyond the approximations implicit in the master equation).

In the examples discussed so far, the parameter value was estimated from the mean value of the measured observable, so the variance of the estimator could be calculated from the error propagation formula (the inverse of the SNR).
If the formula is optimized over the choice of the observable \eqref{eq:fishdef}, estimating from the mean already leads to minimal error.
However, if the observed quantity is fixed (and not necessarily optimal for estimating the parameter), by having access to the full statistics of the measurement results, better precision can sometimes be achieved by choosing a more efficient estimator.
Let $p_\var(i)$ be the probability of obtaining the measurement result $i$, while the true value of the parameter is $\var$.
Then the classical Fisher information (FI) is given as:
\begin{equation}
\label{eq:classicalfish}
    F_{\var}=\sum_i \frac{1}{p_\var(i)}\left(\frac{\partial p_\var(i)}{\partial \var}\right)^2.
\end{equation}
Consider a fixed positive operator-valued measure (POVM) $\{M_i\}_i$, so that $p_\var(i)=\tr(\rho_\var M_i)$.
Then we may define an observable as $\hat O=\sum_i \tilde\var(i)M_i$ and the FI may be seen as the SNR \eqref{eq:fishdef} optimized over $\tilde\var(i)$ with fixed $M_i$.

\subsection{Loss rate estimation}
\label{sec:loss}

We consider again the system governed by the equation \eqref{eq:model} with $H=0$, where the parameter to be estimated is $\Gamma$. The bound \eqref{eq:noisebound} gives:
\begin{equation}
\label{eq:gamma}
    \frac{\F_\Gamma(t)}{t}\leq \frac{\braket{a^\dag a}_{\max} (1+2n_E)}{\Gamma}+\frac{n_E }{\Gamma}.
\end{equation}
For the squeezed displaced state of light  $\mathcal D(\alpha)S(r)\ket{0}$ as an input (in this case squeezing is in the same direction as displacement), we have:
\begin{equation}
\begin{split}
\braket{\hat x(t)}&=2\alpha e^{-\Gamma t/2}\\
\Delta^2 \hat x(t)&=e^{-2r}e^{-\Gamma t}+(1-e^{-\Gamma t})(1+2n_E),
\end{split}
\end{equation}
so
\begin{equation}
\label{eq:gammasnr}
  \frac{S_\Gamma(t)}{t}=\frac{\alpha^2t}{e^{-2r}+(e^{\Gamma t}-1)(1+2n_E)}.
\end{equation}
Note, that the formula for the SNR is exactly the same as the one for frequency estimation. Indeed, while looking only at averages of quadratures, in both cases small changes of the parameter result in the displacement being multiplied by $\alpha^2$.
Therefore, similarly to frequency estimation, for $1/N\ll e^{-2r}\ll \Gamma t \ll 1$ we obtain roughly $S_\Gamma(t)/t\approx N/\Gamma(1+2n_E)$.
Using instead coherent light for a time $t=1/\Gamma$ in each experiment, we loose only a constant factor of order $1/e$.
See also \cite{birchall2020quantum,gianani2021kramers} for more relations between frequency and loss estimation.

While in the limit of small temperature $n_E\to 0$ these results are close to saturating the bound of \eqref{eq:gamma}, they completely fail for large temperature (where, looking at the bound, in principle a larger temperature may help).
One may notice that, unlike $\omega$ in frequency estimation, the changes of parameter $\Gamma$ affect not only the mean of the Gaussian but its covariance matrix as well, so the homodyne detection is no longer the optimal measurement\footnote{The only exception is a coherent state with zero-temperature bath. Then the covariance matrix is not affected by $\Gamma$ and all information is included solely in the mean value of the position.}.

Indeed, to saturate the bound~\eqref{eq:gamma} let us consider a squeezed vacuum state, which is known to be a superposition of the Fock state with an even number of photons.
For a short time, approximately only one photon will be lost or taken from the environment.
As a result, the probability that after small time $tN\Gamma(1+n_E)\ll 1$ the photon number is odd is given by: 
$p_{\t{odd}}\approx\Gamma t\left[(1+2n_E)N+n_E\right]$, see App. \ref{app:loss}.
Using the formula for classical FI~\eqref{eq:classicalfish} we immediately see that the parity measurement saturates~\eqref{eq:gamma}.
Note, that the parity measurement may be easily realized as a quantum nondemolition measurement (QND)~\cite{leghtas2013hardware}, which makes this strategy especially useful in practice.

This problem has been analyzed before in non-adaptive scenarios, both for zero temperature~\cite{monras2007optimal,Adesso2009} and for finite temperature~\cite{wang2020quantum, jonsson2022gaussian}, which converge to our results after taking the appropriate limits.
A different bound appears in~\cite{gagatsos2017bounding}, yet it is looser than ours.
Contrary to all mentioned papers, our bound is guaranteed to remain valid for any adaptive strategy and additional Hamiltonian controls (see \cite{rossi2016enhanced} for example, where non-linearity is added to the Hamiltonian).

\subsection{Temperature estimation $n_E$}
\label{sec:temp}

Here we consider the estimation of the parameter $n_E$, which corresponds to the temperature of the bosonic environment\footnote{Strictly speaking, the two estimation problems are equivalent if $\omega$ is known, since $n_E=1/(e^{\hbar\omega/k_BT}-1)$.}.
Note the difference with equilibrium thermometry, where the system is assumed to be in the Gibbs state.
On the contrary, here we consider dynamical protocols, where unitary operations and measurements can be implemented before reaching equilibrium~\cite{hovhannisyan2021optimal}.
Only such an approach allows for properly including the total time as a resource.
In such formulation, the problem has been intensively discussed for finite-dimensional systems in~\cite{Sekatski2022optimal} and in the Gaussian domain in~\cite{Mirkhalaf2024}.

The parameter $n_E$ to be estimated is a scalar multiplying the Lindblad operators, and 
the bound \eqref{eq:noisebound} gives:
\begin{equation}
\label{eq:temp}
    \frac{\F_{n_E}(t)}{t}\leq \frac{(1+2n_E)\braket{a^\dag a}_{\max}\Gamma}{n_E(1+n_E)}+\frac{\Gamma}{n_E}
\end{equation}
Similarly to the previous case, this bound scales linearly with the number of photons.
Note, however, that here using a coherent state offers no advantage over using a vacuum---indeed, the output state is the same as for the vacuum, only displaced by an $n_E$-independent factor.
Therefore, for obtaining the optimal scaling with $N$ it is necessary to use a nonclassical state of light.

As an exemplary strategy for saturating the bound, consider preparing the initial state as the Fock state $\ket{N}$ and performing photon counting after a small time.
For $tN\Gamma(1+n_E)\ll 1$, we have approximately (see App. \ref{app:temperature} for exact formulas):
 \begin{alignat}{2}
        &p(N)&&=1-tN\Gamma(1+n_E)-t(N+1)\Gamma n_E\\
         &p(N+1)&&=t(N+1)\Gamma n_E\\
          &p(N-1)&&=tN\Gamma(1+n_E).
    \end{alignat}
The corresponding classical FI reads:
\begin{equation}
    F_{n_E}\overset{Nt\Gamma\ll 1}{\approx}\frac{\Gamma N t(1+2n_E)}{n_E(1+n_E)}+\frac{\Gamma t}{n_E},
\end{equation}
so it saturates the bound.
This strategy works similarly well also for the previous problem of estimating the loss rate $\Gamma$.

Note that, unlike loss estimation, the parity measurement would not be sufficient for optimal performance.
This happens because for temperature estimation it is crucial to distinguish between when a photon is lost and when it is taken from the environment; see App. \ref{app:loss} for the mathematical explanation.
A nondemolition photon counting measurement is more challenging to implement than a parity measurement.
However, if one were able to perform it, it would be very useful for fast-prepare-and-measure protocols, as a post-measurement outcome would be a Fock state, which is ready for another iteration.
See ~\cite{dixit2021searching,agrawal2023stimulated} for current progress in the experimental application of a similar strategy in analogous problems. 

This is also why using a single-mode squeezed state does not saturate the bound (unlike loss rate estimation). Indeed, even if it gives a QFI with a linear scaling in $N$~\cite{Mirkhalaf2024}, it is far from being optimal in the limit of small temperature.
The problem can be addressed by using a two-mode squeezed vacuum as the input state, where the second mode serves as a noiseless ancilla~\cite{shi2023ultimate}, saturating \eqref{eq:temp} in the limit of short single-iteration times.

To better understand our result, it is worth comparing it with previously known bounds.
The problem has been discussed in the literature, typically with the assumption that after preparing the state, it interacts freely with the bath for some fixed time $t$, after which a single measurement is performed. 

In~\cite{wang2020quantum,shi2023ultimate} another bound 
based on channel purification has been derived, which for small $t$ correspond with \eqref{eq:temp}, see App. \ref{app:temperature} for details.
Another interesting bound is the one from ~\cite{Pirandola2017}, stating that:
\begin{equation}
\label{eq:termsing}
    \F^{\t{single shot}}_{n_E}\leq \frac{1}{n_E(1+n_E)},
\end{equation}
and does not depend on $t,N,\Gamma$ at all (while in practice the number of photons $N$ needed for saturating this bound increases
with the inverse of $t\Gamma$).
This can be intuitively understood because, for a very long time, the system fully thermalizes (while the time of thermalization depends on $N,\Gamma$).

\begin{figure}[t!]
  \includegraphics[width=0.48
\textwidth]{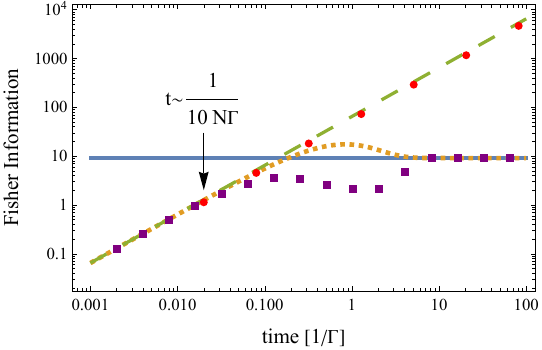}
\centering
\caption{Bounds for the Fisher information and the classical Fisher information for specific strategies, both plotted as functions of time.
Results are shown for $N=5$ and $n_E=0.1$.
The blue solid line \eqref{eq:termsing} and yellow dotted line \eqref{eq:shi} correspond to bounds from the literature, derived for passive strategies (no additional action is performed during time $t$ and the measurement is performed after this time).
Purple squares are the classical Fisher information for passive evolution of the Fock input state~\eqref{eq:clasfish}.
For long evolution time, the state becomes fully thermalized and the FI converges to the bound~\eqref{eq:termsing}.
The green dashed line~\eqref{eq:temp} is a general bound for any strategy using total time $t$ (including measurements between and/or any kind of adaptivity).
Red dots represent an exemplary strategy saturating this bound, based on the fast prepare-and-measure protocol with Fock states, with single-sensing time $t\sim 1/10N\Gamma$, without the necessity of any adaptivity. 
}
\label{fig:temp}
\end{figure}

More formally, the bound may be derived using the concept of quantum simulation~\cite{matsumoto2010metric,kolodynski2014precision}.
The free evolution of any input state of the cavity interacting with the bath may be equivalently obtained by mixing this state with a thermal state via beam-splitter with transmittance $\kappa=e^{-t\Gamma}$.
As the beamsplitter itself does not depend on $n_E$, the QFI of the final state of the system cannot be bigger than the QFI of the initial global state of both systems and environments, which is (due to the product structure of initial state) simply the QFI of the thermal state.
This allows for bounding the problem of channel estimation by the simpler problem of state estimation.

Note, that the bound \eqref{eq:termsing} does not contradict our results, but rather explains the origin of the advantage of using nonclassical states.
Using more photons cannot overcome the bound \eqref{eq:termsing}, but it may allow for saturating it in a shorter time.
As a consequence, using higher-energy Fock states more measurement repetitions may be performed in the same total time.
Compared to previous results, our bound shows that this result cannot be overcome even by using an adaptive strategy.
The relation of our result to the existing bounds is summarized in~\autoref{fig:temp}.

A comparison of \eqref{eq:termsing} with \eqref{eq:temp} shows that, for optimal performance,   the time of a single repetition cannot be longer than $t\sim 1/\Gamma N$. One should be careful with taking the limit $N\to\infty$, as then one can reach time scales so small, that the Lindblad equation is not valid any more.

Note, that the problem of temperature estimation is closely related to dark-matter sensing, where light displacement with a completely random direction is to be detected~\cite{shi2023ultimate,shi2024quantum}, see App. \ref{app:axion} for details.

Here the impact of temperature is modeled only via exchanging the photons with the environment.
As pointed out in~\cite{Sekatski2022optimal} the environment also causes a Lamb shift, which generally depends on the temperature and thus significantly affects the bound on the QFI (especially for small temperatures).
However, it is not clear how to saturate the contribution coming from this term with a practical strategy~\cite{Sekatski2022optimal}.
Interestingly, the situation is different for bosonic systems, for which a microscopic derivation of the master equation shows that the Lamb shift does not depend on the temperature~\cite{Mirkhalaf2024}.

\section{Conclusions}

We investigated the interplay between time and energy in quantum sensing with infinite-dimensional systems.
We considered quantum estimation protocols for all the parameters of a minimal model for a quantum resonator in the open-quantum system setting.
In particular, we considered noisy quantum estimation of frequency, squeezing, loss rate, and bath temperature.
In all cases, we treat both the number of photons and sensing time as a resource. Comparing fundamental bounds with simple protocols, we found that for frequency and displacement estimation almost optimal performance may be obtained by using coherent light.
On the contrary, in noise estimation optimal performance requires using of nonclassical states of light. Especially, in temperature estimation, it allows for a better scaling with the number of photons.

In the analysis, we did not include the system preparation time and measurement time, which we call jointly $t_{p,m}$.
If one were to include them, one should optimize $I(t)/(t+t_{p,m})$ instead of $I(t)/t$. However, regarding the analysis performed, the inclusion of these times would work in favor of the protocols with classical light over the ones with squeezed light (since for them the time of a single measurement is longer, so $I(t)/(t+t_{p,m})$ will suffer less from adding $t_{p,m}$. It is worth mentioning, that in this context the critical metrology may outperform both of them, as discussed in \cite{alushi2024optimality}. Moreover, the preparation time becomes negligible in the protocol, where continuous measurement is applied for the steady state of driven cavity. Also in the case of loss estimation, this problem may be completely omitted by using non-demolition parity measurement.

In this paper, we performed an analysis based on the QFI and SNR.
Even though these tools sometime do not give the full picture when reaching the Heisenberg limit~\cite{hayashi2011,Giovannetti2012beyond,hall2012universality,Pezze2018,Gorecki2020pi}, in the cases discussed, the asymptotic scaling of precision with time was only linear, so the results we obtained are fully trustable~\cite{gill2000state,kahn2008local,Haah2017sample,yang2019attaining}.
In addition, for example, the temperature estimation problem has also been discussed in the literature using Bayesian estimation theory (for the finite-dimensional case), with qualitatively similar results~\cite{bavaresco2023designing}.

In conclusion, we have identified optimal strategies for noisy parameter estimation in a paradigmatic infinite-dimensional quantum model, of high experimental relevance.
In doing so, we have also shown that fundamental bounds for arbitrary possible strategies, including adaptivity and ancillae, can be saturated by simple passive protocols in most cases.
We believe these results show that deriving fundamental bounds for quantum metrology can have a concrete impact on the design of optimal protocols and practical sensing devices.
In particular, we have also highlighted the usefulness of these bounds not only for Hamiltonian parameters but also for noise parameters.
Since such methods are universal, we expect that they will be relevant in other scenarios and physical systems, including atoms or other quantum nonlinearities, e.g. to study the estimation of two-photon absorption~\cite{sanchezmunozQuantumMetrologyTwophoton2021,panahiyanTwophotonabsorptionMeasurementsPresence2022,Karsa2024a}.

\section*{Acknowledgements}
W.G. acknowledges financial support from the
U.S. Department of Energy, Office of Science, National
Quantum Information Science Research Centers, Superconducting Quantum Materials and Systems Center
(SQMS) under the contract No. DE-AC02-07CH11359.
F.A. acknowledges support from Marie Skłodowska-Curie Action EUHORIZON-MSCA-2021PF-01 (project \mbox{QECANM}, grant no. 101068347).
S.F. acknowledges financial support from the foundation Compagnia di San Paolo, grant vEIcolo no. 121319, and from PNRR MUR project PE0000023-NQSTI financed by the European Union – Next Generation EU.
R.D. acknowledges support from the Academy of Finland, grants no. 353832 and 349199.
L.M.  acknowledges support from the PNRR MUR Project PE0000023-NQSTI and from the National Research Centre for HPC, Big Data and Quantum Computing, PNRR MUR Project CN0000013-ICSC and from 765
the PRIN2022374 CUP Grant No. 2022RATBS4.

\appendix

\section{QFI as maximized SNR}
\label{app:QFI}
Here we prove \eqref{eq:fishdef}, i.e.:
\begin{equation}
\F_\var=\max_{\hat O}S,\quad \t{where}\quad S=\frac{|\partial_\var \braket{\hat O}|^2}{\Delta^2\hat O}.
\end{equation}
By definition, the QFI is equal to:
\begin{equation}
\F_\var =\tr(\rho_\var \Lambda^2),
\end{equation}
where the symmetric logarithmic derivative $\Lambda$ is a solution of:
\begin{equation}
\label{eq:SLD}
\frac{1}{2}(\rho_\var\Lambda+\Lambda\rho_\var)=\frac{\partial \rho_\var}{\partial\var}.
\end{equation}
We now move to the maximization of $S$ over $\hat O$.
Note that the formula for $S$ is invariant under rescaling $\hat O$ by a constant, as the square of the rescaling factor appears in both the numerator and the denominator.
Therefore, without loss of generality we may impose the condition $\tr\left(\frac{\partial\rho_\var}{\partial\var}\hat O\right)=1$ to simplify further reasoning.
Then the problem reduces to minimization of $\Delta^2\hat O$.
Moreover, minimization of $\Delta^2\hat O$ gives the same results as minimization of $\tr(\rho_\var\hat O^2)$ (as one may always replace $\hat O\to \hat O-\tr(\rho_\var \hat O)\openone$ without violating the imposed condition). Therefore, the problem may be formulated as follows:
\begin{equation}
\min_{\hat O}\tr(\rho_\var \hat O^2),\quad \t{with}\quad \tr\left(\frac{\partial\rho_\var}{\partial\var}\hat O\right)=1.
\end{equation}
To solve it, we use the method of Lagrange multipliers.
Consider small change $\hat O\to \hat O+\delta \hat X$, with $\delta$ being a small scalar and $\hat X$ being arbitrary hermitian matrix.
The solution needs to satisfy:
\begin{multline}
\tr(\rho_\var (\hat O+\delta \hat X)^2)-\tr(\rho_\var \hat O^2)=\\\lambda \tr\left(\frac{\partial\rho_\var}{\partial\var}(\hat O+\delta\hat X)\right)+\mathcal O(\delta^2),
\end{multline}
with $\lambda$ being a Lagrange multiplier.
Since we consider arbitrarily small $\delta$, terms including $\delta^2$ become irrelevant.
The equation needs to be satisfied for any $\hat X$, thus we have:
\begin{equation}
\label{eq:O}
\rho_\var \hat O+\hat O\rho_\var=\lambda \frac{\partial\rho_\var}{\partial\var}
\end{equation}
One can see that the above equality is the same as \eqref{eq:SLD}, up to a constant multiplicative factor $\lambda$, so $\hat O_{\t{opt}}= \frac{\lambda}{2}\Lambda$.
Taking into account the condition $\tr\left(\frac{\partial\rho_\var}{\partial\var}\hat O_{\t{opt}}\right)=1$, and using the equality \eqref{eq:SLD} we find that $\frac{\lambda}{2}=\left[\tr(\rho_\var\Lambda^2)\right]^{-1}$, so indeed:
\begin{equation}
\max_{\hat O}\frac{|\partial_\var \braket{\hat O}|^2}{\Delta^2\hat O}=\tr(\rho_\var \Lambda^2).
\end{equation}

\section{General version of the bound}
\label{app:general}
In the most general case, one may consider the situation when the parameter is encoded in both Hamiltonian and dissipative parts simultaneously. All bounds mentioned in the main text may be seen as the consequence of the bound for the derivative of QFI over time, being a generalization of the results from \cite{kurdzialek2022using,das2024universal} for unbounded operators:
\begin{theorem}
\label{thm:continous}
For the evolution described in \eqref{eq:lind}, derivative of the QFI of evolving state $\rho(t,\var)$ over time is fundamentally bounded by
\begin{equation}
\label{eq:thm1}
   \frac{d \F_\varphi(t)}{dt} \leq 4\min_{h} \left(\braket{\mathfrak{a}(h)}+\sqrt{\braket{\mathfrak{b}(h)^2}\F_\varphi(t)}\right),
\end{equation}
where the optimization variable
\begin{equation}
h = 
    \begin{pmatrix}
        \begin{array}{c|c}
             h_{00} & \vec{h}^\dagger \\
             \hline
             \vec{h} & \mathfrak{h} \\    
        \end{array}
    \end{pmatrix}
\end{equation}
is an $(J+1) \times (J+1)$ Hermitian matrix with a block structure, where $\mathfrak{h} \in \mathbb{C}^{J \times J}_{\t{H}}$ is an $J\times J$ hermitian matrix, $\vec{h} \in \mathbb{C}^J$ is a complex vector of length $J$, $h_{00} \in \mathbb{R}$, $\braket{\cdot}$ is an expectation value, while $\mathfrak{a}(h)$, $\mathfrak{b}(h)$ are operators defined as:
\begin{equation}
\label{eq:aoperator1}
\begin{split}
\mathfrak{a}(h) & = {\left(i\dot{\vec{L}}+ \mathfrak{h} \vec{L} + \vec{h} \openone \right)}^\dagger \left(i\dot{\vec{L}}+ \mathfrak{h} \vec{L} + \vec{h} \openone\right),\\
\mathfrak{b}(h) & =\dot{H} - \frac{i}{2} \left( \dot{\vec{L}}^\dagger \vec{L} - \vec{L}^\dagger \dot{\vec{L}} \right)  + h_{00}\openone+\vec{L}^\dag \vec{h}+\vec{h}^\dag\vec{L}+\vec{L}^\dag \mathfrak h \vec{L}.
\end{split}
\end{equation}
Here $\dot{X}=\partial_\var X$ denotes derivative over parameter, $\vec{L}=[L_1,...,L_J]^T$ is a vector of Linblad operators.
\end{theorem}

A similar (and slightly stronger) bound has been derived with alternative methods in ~\cite{wan2022bounds}. 

Note that the analysis of unbounded operators in continuous-variable quantum systems in the context of quantum metrology requires special attention, as the moments of observables can diverge in such cases.
Thus, in principle the expectation values $\braket{\mathfrak{a}(h)},\braket{\mathfrak{b}(h)}$ appearing in the final bound may also diverge.
This may happen especially in the case of noiseless evolution, when Heisenebrg scaling occurs.
For such a case indeed one may obtain an arbitrarily large QFI, but a reasonable analysis taking into account also the saturability of the Cram\'er-Rao inequality requires using more advance methods, which finally leads to a finite precision ~\cite{hayashi2011,Giovannetti2012beyond,hall2012universality,Gorecki2020pi}.
This, however, lies beyond the scope of the current paper, as this issue does not arise in the examples discussed here.

To prove \eqref{eq:thm1}, we start with a stronger version of the results from \cite{kurdzialek2022using}, generalized for unbounded operators:

\begin{lemma}
Consider the state $\rho_\var$ depending on the unknown parameter $\var$, which is next treated as an input for the channel depended on the same parameter $\Lambda_\var[\cdot]=\sum_{k=0}^JK_k(\var)\cdot K^\dagger_k(\var)$.
Then the difference between the QFI of the the final state $\Lambda_\var[\rho_\var]$ and $\rho_\var$ is bounded by:
\begin{equation}
\label{eq:krauss}
\F_\var[\Lambda_\var[\rho_\var]]-\F_\var[\rho_\var]\leq 4\left(\braket{\alpha}+\sqrt{\braket{\beta^2}\F_\var[\rho_\var]}\right)
\end{equation}
where $\alpha=\dot{\vec{K}}^\dagger\dot{\vec{K}}$, $\beta=i\vec{K}^\dagger\dot{\vec{K}}$.\end{lemma}

\textit{Proof.} See \cite{gorecki2024tba}.$\square$

Having that, we can prove \autoref{thm:continous}.

\textit{Proof of \autoref{thm:continous}}.
The evolution governed by a Lindblad equation, for small time $\Delta t$ may be approximated up to terms $\mathcal O(\Delta t^{3/2})$ by acting with the Kraus operators \cite{demkowicz2017adaptive}:
\begin{align}
      K_0 &= \openone - \left( i H + \frac{1}{2} \vec{L}^\dag \vec{L} \right) \Delta t+\mathcal O(\Delta t^{3/2}),\\
    \vec{K} &= \vec{L} \sqrt{\Delta t}+\mathcal O(\Delta t). 
\end{align}
After substituting into \eqref{eq:krauss}, dividing both sides by $\Delta t$ and taking the limit $\Delta t\to 0$, one obtains:
\begin{equation}
\label{eq:simple}
   \frac{d \F_\varphi(t)}{dt} \leq 4 \left(\braket{\mathfrak{a}}+\sqrt{\braket{\mathfrak{b}^2}\F_\varphi(t)}\right),
\end{equation}
where $\aaa=\dot{\vec{L}}^\dagger \dot{\vec{L}}$, $\bbb=\dot H-\frac{i}{2}(\dot{\vec{L}}^\dag \vec{L}-\vec{L}^\dag \dot{\vec{L}})$.

Next, note that both the Kraus representation and Lindblad operators are not unique.
To make the bound tighter, in \cite{zhou2018achieving,demkowicz2017adaptive} an optimization over Kraus representation has been performed.
Here we perform an analogous optimization over the Lindblad operators, leading to the same result.
Indeed, after the following substitution:
\begin{equation}
\begin{split}
 \vec{L}^\prime &=   e^{-i\mathfrak h \var}\vec{L} + \var \vec{h}\openone,\\
 H^\prime &= H + 
 \frac{\var}{2i}\left[\vec{h}^\dagger \vec{L} - \vec{L}^\dagger \vec{h}\right]+\var h_{00}\openone,
 \end{split}
\end{equation}
the evolution remains unchanged, so one may optimize the bound \eqref{eq:simple} over $h$ to get \eqref{eq:thm1}.

\section{Calculations of bound for discussed examples}
\label{app:calculations}

\textbf{Frequency estimation.}
For frequency estimation in \autoref{sec:freq}, only the diagonal elements of $h$ are relevant to impose the condition $\bbb(h)=0$.
Setting all off-diagonal elements equal to $0$ we have:
\begin{equation}
\begin{split}
\bbb(h)&=a^\dag a+h_{00}+h_{11}\Gamma(1+n_E)a^\dag a+h_{22}\Gamma n_E a a^\dag,\\
\aaa(h)&=(h_{11})^2\Gamma(1+n_E)a^\dag a+(h_{22})^2\Gamma n_Ea a^\dag,
\end{split}
\end{equation}
so the problem is formulated as:
\begin{equation}
\begin{split}
\min_{h_{00},h_{11},h_{22}}&
(h_{11})^2\Gamma(1+n_E)\braket{a^\dag a}+(h_{22})^2\Gamma n_E(\braket{a^\dag a}+1),\\
&\t{with}
\quad 
\begin{cases}
    1+h_{11}\Gamma(1+n_E)+h_{22}\Gamma n_E=0,\\
    h_{00}+h_{22}\Gamma n_E=0,
\end{cases}
\end{split}
\end{equation}
which after direct optimization gives:
\begin{equation}
\braket{\aaa(h)}=\frac{\braket{a^\dag a}}{\Gamma(1+2n_E-\frac{n_E}{\braket{a^\dag a}+1})}\overset{\braket{a^\dag a}\gg 1}{\approx} \frac{\braket{a^\dag a}}{\Gamma(1+2n_E)}.
\end{equation}

\textbf{Displacement estimation.}
For displacement estimation in \autoref{sec:disp}, only the $\vec{h}$ elements of $h$ are relevant to impose the condition $\bbb(h)=0$:
\begin{equation}
\begin{split}
\bbb(h)=&ia^\dag-ia+h_{01} \sqrt{\Gamma(1+n_E)}a+h_{02} \sqrt{\Gamma n_E}a^\dagger,\\
&+
h_{10}\sqrt{\Gamma(1+n_E)}a^\dagger +h_{20} \sqrt{\Gamma n_E}a\\
\aaa(h)=&(|h_{01}|^2+|h_{02}|^2)\openone,
\end{split}
\end{equation}
so:
\begin{equation}
\begin{split}
\min_{h_{01},h_{02}}&
|h_{01}|^2+|h_{02}|^2,\\
&\t{with}
\quad 
\begin{cases}
    i+h_{01}^*\sqrt{\Gamma(1+n_E)}+h_{02}\sqrt{\Gamma n_E}=0,\\
    -i+h_{01}\sqrt{\Gamma(1+n_E)}+h_{02}^*\sqrt{\Gamma n_E}=0,
\end{cases}
\end{split}
\end{equation}
which gives
\begin{equation}
   \braket{\aaa(h)}=\frac{1}{\Gamma (1+2n_E)}.
\end{equation}

\textbf{Squeezing estimation.}
For displacement estimation in \autoref{sec:squeezing}, only the off-diagonal element of $\mathfrak h$ is relevant to impose the condition $\bbb(h)=0$:
\begin{equation}
\begin{split}
\bbb(h)=&a^2+a^{\dagger 2}+h_{21}a^2 \Gamma \sqrt{n_E(1+n_E)}\\
&+h_{12}a^{\dagger 2}\Gamma \sqrt{n_E(1+n_E)}\\
\aaa(h)=&|h_{12}|^2(2a^\dagger a+1)\Gamma \sqrt{n_E(1+n_E)}.
\end{split}
\end{equation}
The unique solution giving $\bbb(h)=0$ is therefore $h_{12}=-(\Gamma \sqrt{n_E(1+n_E)})^{-1}$, which gives:
\begin{equation}
\braket{\aaa(h)}= \frac{2\braket{a^\dag a}+1}{\Gamma \sqrt{n_E(1+n_E)}}.
\end{equation}

\section{Frequency estimation---continuous measurement of the output field}
\label{app:cont}
We consider a single-mode resonator $a$ of frequency $\omega$, coupled to a dissipative environment $b_{\t{in}}$ with constant $\Gamma$,  and coupled to input-output modes with constant $\gamma$, as sketched in~\figref{fig:cav}.
\begin{equation}
    \dot{a}(t)=-i\omega a(t)-\frac{1}{2}(\Gamma+\gamma) a(t)+\sqrt{\Gamma}b_{\t{in}}(t)+\sqrt{\gamma}c_{\t{in}}(t).
\end{equation}
The environment modes obey the bosonic commutation relations $[c_{\t{in}}(t), c_{\t{in}}^\dag(t')]= \delta(t-t')$, and analogously for $b_{\t{in}}(t)$.
We assume that the dissipative environment modes $b_{\t{in}}(t)$ are in the vacuum, while $c_{\t{in}}(t)$ is given by monochromatic coherent light with frequency $\omega_0$ and photon flux $|\alpha_f|^2$ (where $|\alpha_f|^2$ has dimension $[1/s]$), so that $\braket{c^\dagger_{\t{in}}(t)c_{\t{in}}(t')}=|\alpha_f|^2 e^{-i(t'-t)\omega_0}$.
The formal solution to this equation is:
\begin{multline}
    a(t)=a(0)e^{-(i\omega+(\Gamma+\gamma)/2)t}\\
    +e^{-(i\omega+(\Gamma+\gamma)/2)t} \int_0^t e^{(i\omega+(\Gamma+\gamma)/2)\tau}(\sqrt{\Gamma}b_{\t{in}}(\tau)+\sqrt{\gamma}c_{\t{in}}(\tau))d\tau.
\end{multline}
From direct calculation, the mean number of photons for the steady state is given as:
\begin{equation}    N_{ss}:=\lim_{t\to\infty}\braket{a^\dagger(t)a(t)}=\frac{\gamma|\alpha_f|^2}{(\frac{\Gamma+\gamma}{2})^2+(\omega-\omega_0)^2},
\end{equation}
where the steady state is obtained for $t\gg 1/(\Gamma+\gamma)$. The output flux mode is given by the input-output relation:
\begin{equation}
    c_{\t{out}}(t)=\sqrt{\gamma}a(t)-c_{\t{in}}(t).
\end{equation}
The power of the output mode is:
\begin{equation}    n_{\t{out}}:=\braket{c_{\t{out}}(t)^\dagger c_{\t{out}}(t)}=|\alpha_f|^2-\frac{\Gamma\gamma|\alpha_f|^2}{(\frac{\Gamma+\gamma}{2})^2+(\omega-\omega_0)^2}.
\end{equation}
These stay consistent with the observation that the power of mode $b_{\t{out}}(t)$ is trivially $\Gamma N_{ss}$, and the total number of photons is conserved.

Now, we want to estimate $\omega$ from the number of photons counted in total time $t$, 
\begin{equation}
    N_{\t{out}}^t=\int_0^t n_{\t{out}}(\tau)d\tau.
\end{equation}
The signal-to-noise ratio is given as:
\begin{equation}
    S_{\omega}(t)=\frac{|\partial_\omega N_{\t{out}}^t|^2}{\Delta^2 N_{\t{out}}^t},
\end{equation}
where
\begin{equation}
    |\partial_\omega N_{\t{out}}^t|^2=\left|\partial_{\omega}\int_{0}^t\braket{c^\dagger_{\t{out}}(\tau)c_{\t{out}}(\tau)} d\tau \right|^2
\end{equation}
    and
 \begin{multline}
     \Delta^2 N_{\t{out}}^t=
 \int_0^td\tau\int_0^td\tau' \Big[\braket{c^\dagger_{\t{out}}(\tau)c_{\t{out}}(\tau)c^\dagger_{\t{out}}(\tau')c_{\t{out}}(\tau')}\\
 -\braket{c^\dagger_{\t{out}}(\tau)c_{\t{out}}(\tau)}\braket{c^\dagger_{\t{out}}(\tau')c_{\t{out}}(\tau')}\Big].
\end{multline}
Note that:
\begin{multline}
    \braket{c^\dagger_{\t{out}}(\tau)c_{\t{out}}(\tau)c^\dagger_{\t{out}}(\tau')c_{\t{out}}(\tau')}\\=\braket{c^\dagger_{\t{out}}(\tau)c^\dagger_{\t{out}}(\tau')c_{\t{out}}(\tau)c_{\t{out}}(\tau')}+\delta(\tau-\tau')\braket{c^\dagger_{\t{out}}(\tau)c_{\t{out}}(\tau)}
\end{multline}
and, since $c_{\t{in}}(\tau)$ is coherent, 
\begin{multline}
   \braket{c^\dagger_{\t{out}}(\tau)c^\dagger_{\t{out}}(\tau')c_{\t{out}}(\tau)c_{\t{out}}(\tau')}=\\
   \braket{c_{\t{out}}^\dagger(\tau)c_{\t{out}}(\tau)}\braket{c_{\t{out}}^\dagger(\tau')c_{\t{out}}(\tau')}
\end{multline}
so the SNR simplifies to:
\begin{equation}
   S_{\omega}(t)=\frac{t|\partial_\omega n_{\t{out}}|^2}{n_{\t{out}}}.
\end{equation}
The full formula is:
\begin{equation}
   S_{\omega}(t)=\frac{1024|\alpha|^2(\omega-\omega_0)^2\gamma^2\Gamma^2T}{(4(\omega-\omega_0)^2+(\Gamma-\gamma)^2)(4(\omega-\omega_0)^2+(\Gamma+\gamma)^2)^3}.
\end{equation}
For the specific value $\gamma=\Gamma$, the output power simplifies to:
\begin{equation}
    n_{\t{out}}=\frac{(\omega-\omega_0)^2|\alpha|^2}{\Gamma^2+(\omega-\omega_0)^2}.
\end{equation}
Note that exactly at resonance $\omega=\omega_0$, the power of the output mode is zero.
This is because the light from the coherent source and the light from the cavity interfere destructively.
Simultaneously, the power of $b_{\t{out}}(t)$ becomes equal to the power of a laser.
Then, since
\begin{equation}
   N_{ss}\overset{\gamma=\Gamma}{=}\frac{\Gamma|\alpha|^2}{\Gamma^2+(\omega-\omega_0)^2}\overset{\omega\approx \omega_0}{\to}\frac{|\alpha|^2}{\Gamma}
\end{equation}
we have
\begin{equation}
   S_{\omega}(t)=\frac{4|\alpha|^2\Gamma^4 T}{(\Gamma^2+(\omega-\omega_0)^2)^3}\overset{\omega\approx \omega_0}{\to}\frac{4N_{ss}T}{\Gamma},
\end{equation}
which saturates the bound.

Note that the optimal point $\omega\approx \omega_0$ is not the one for which the mean number of photons $N_{ss}$ has a strong dependence on $\omega$, but the one where $N_{ss}$ attains its maximum, and, because of that, $n_{\t{out}}$ goes to zero---which effectively reduces the denominator in the SNR.
In the above reasoning, two limits have been taken: $\lim_{\omega\to\omega_0}\lim_{\gamma\to\Gamma}$ and the order of taking the limit matters.
In practice, it means that the final formula works in the regime where:
\begin{equation}
    \Gamma^2\gg (\omega-\omega_0)^2\gg (\Gamma-\gamma)^2
\end{equation}
so, depending on how accurately one can set $\gamma$ close to $\Gamma$, one should intentionally increase the difference between $\omega$ and $\omega_0$.

One thing to notice is that both the signal and the noise in the SNR are going to $0$.
While this is a well-defined mathematical limit, it means that the SNR is extremely sensitive to any kind of additional noise appearing in practice (for example, if the photon counter makes false-positive detection from time to time, it completely destroys the protocol).
One may always choose $\omega_0\approx \omega+\Gamma$ to make it stable (but then SNR is reduced by a constant factor).

\section{Squeezing estimation---QEC protocol}
\label{app:squeezing}
Assume that the states of the system $\mathcal H_S$ may be entangled with a noiseless ancilla $\mathcal H_A$.
An effective quantum error correction protocol for estimating a parameter of the Hamiltonian $H$ may be applied iff there exists a proper code space $\mathcal H_C\subset \mathcal H_S\otimes \mathcal H_A$ satisfying (for simplicity we consider the situation when there is only one Linbdlad operator, which is our case)~\cite{zhou2018achieving}:
\begin{equation}
\label{eq:cond}
\begin{split}
&\Pi_C (\dot H\otimes \openone_A)\Pi_C\not\propto \openone\\
&\Pi_C (L\otimes \openone_A)\Pi_C=\lambda\openone\\
&\Pi_C (L^\dagger L\otimes \openone_A)\Pi_C= \mu\openone,
\end{split}
\end{equation}
where $\Pi_C$ is a projection onto code space $\mathcal H_C$, $\dot H$ denotes the derivative of Hamiltonian over the estimated parameter, while $\lambda$ and $\mu$ are scalars.
Once the above are satisfied, after applying a proper QEC, one obtains a unitary evolution $U=\exp(-it \Pi_C (H\otimes\openone_A) \Pi_C)$ on this subspace.

The first condition guarantees that the Hamiltonian acts non-trivially inside the code space, so there appears the signal to be measured.
The second condition guarantees that the noise does not act inside of the code space (it may only take the state outside of it).
The last condition is introduced to guarantee that the orthogonal states of the code space are still perfectly distinguishable after the action of the noise, which is necessary for performing QEC.

To satisfy conditions \eqref{eq:cond} for the squeezing estimation problem in \autoref{sec:squeezing}, we choose the subspace spanned by two states:
\begin{equation}
\begin{split}
\ket{c_0}&=\tfrac{1}{\sqrt{2}}(\ket{N-2}+\ket{N})\otimes \ket{0}_A\\
\ket{c_1}&=\tfrac{1}{\sqrt{2}}(\ket{N-2}-\ket{N})\otimes \ket{1}_A,
\end{split}
\end{equation}
where $\ket{0}_A,\ket{1}_A$ are mutually orthogonal states of ancilla.

One can easily check that $\Pi_{\mathcal C}(\dot H\otimes \openone_A)\Pi_{\mathcal C}=\sqrt{N(N-2)}\sigma_z$, which would lead to $F=4N(N-2)t^2$.

\section{Squeezing estimation---cat state protocol}
\label{app:catstate}

In this appendix we show that for short time, satisfying $t \epsilon\ll 1/\sqrt{N+2}$, the action of the Hamiltonian $\epsilon(a^{\dagger 2}+a^2)$ on cat states may be well approximated by:
\begin{equation}
\begin{split}
    (a^{\dagger 2}+a^2)\ket{\mathcal C_{\alpha}^\pm}&\approx 2\t{Re}(\alpha^2)\ket{\mathcal C_{\alpha}^\pm},\\
    (a^{\dagger 2}+a^2)\ket{\mathcal C_{i\alpha}^\pm}&\approx-2\t{Re}(\alpha^2)\ket{\mathcal C_{\alpha}^\pm},
\end{split}
\end{equation}
i.e., the correction to this approximation, even accumulated over this time, is still negligible.


A projection of $a^\dag \ket{\alpha}$ on $\ket{\alpha}$ gives:
\begin{equation}
|\braket{\alpha|a^\dagger|\alpha}|^2=|\alpha|^2
\end{equation}
while the norm of $a^\dag \ket{\alpha}$ is
\begin{equation}
|\braket{\alpha|a a^\dagger|\alpha}|^2=|\alpha|^2+1,
\end{equation}
from which:
\begin{equation}
a^\dag \ket{\alpha}=\alpha^*\ket{\alpha}+\ket{\alpha^\perp}
\end{equation}
where $\ket{\alpha^\perp}$ is normalized and orthogonal to $\ket{\alpha}$.

Similar reasoning may be performed for $a^{\dag 2} \ket{\alpha}$ and $\alpha^{* 2}\ket{\alpha}$.
The projection of $a^{\dag 2} \ket{\alpha}$ on $\ket{\alpha}$ gives:
\begin{equation}
|\braket{\alpha|a^{\dagger 2}|\alpha}|^2=|\alpha|^4
\end{equation}
while the norm of $a^{\dag 2} \ket{\alpha}$ is:
\begin{equation}
|\braket{\alpha|a^2  a^{\dagger 2}|\alpha}|^2=|\alpha|^4+4|\alpha|^2+2
\end{equation}
so
\begin{equation}
(a^\dag)^2\ket{\alpha}=(\alpha^*)^2\ket{\alpha}+\sqrt{4|\alpha|^2+2}\ket{\alpha'^{\perp}}
\end{equation}
Therefore we have:
\begin{equation}
\begin{split}
    \|(a^{\dagger 2}+a^2)\ket{\mathcal C_{\alpha}^\pm}-2\t{Re}(\alpha^2)\ket{\mathcal C_{\alpha}^\pm}\|&=\sqrt{4|\alpha|^2+2},\\
    \|(a^{\dagger 2}+a^2)\ket{\mathcal C_{i\alpha}^\pm}-2\t{Re}(\alpha^2)\ket{\mathcal C_{\alpha}^\pm}\|&=\sqrt{4|\alpha|^2+2}.
\end{split}
\end{equation}

Now we consider how the correction may accumulate over time.
We start from the state $\ket{\psi(0)}$ and analyze its evolution govern by Hamiltonian $H$, namely $\ket{\psi(t)}=e^{-itH}\ket{\psi(0)}$.
Let $H_C$ be an approximation of $H$.
The question is: how well can the Hamiltonian $H$ be approximated by $H_C$?
More precisely, having $\ket{\psi_C(t)}=e^{-itH_C}\ket{\psi(0)}$, what will be the bound for $\|\ket{\psi(t)}-\ket{\psi_C(t)}\|$?
By direct calculation:
\begin{multline}
 \frac{d}{dt}\left[(\bra{\psi(t)}-\bra{\psi_C(t)})(\ket{\psi(t)}-\ket{\psi_C(t)})\right]=\\
   2\t{Im}\braket{\psi(t)|(H-H_C)|\psi_C(t)}\leq 2\|(H-H_C)\ket{\psi_C(t)}\|,
\end{multline}
so 
\begin{equation}
\|\ket{\psi(t)}-\ket{\psi_C(t)}\|^2\leq 2t \max_{t'\in[0,t]}\|(H-H_C)\ket{\psi_C(t')}\|.
\end{equation}
In our case:
\begin{equation}
\max_{t'\in[0,t]}\|(H-H_C)\ket{\psi_C(t')}\|\leq \epsilon\sqrt{4|\alpha|^2+2},
\end{equation}
so
\begin{equation}
\|\ket{\psi(t)}-\ket{\psi_C(t)}\|\leq t\epsilon\sqrt{4N+2}.
\end{equation}

\section{Loss rate estimation}

\label{app:loss}
Here we derive the formula for the classical FI with respect to $\Gamma$ for the parity counting measurement, applied to the squeezed vacuum state after a short evolution $(N+1)\Gamma t(n_E+1)\ll 1$.
In particular, we clarify why it contains as much information as exact photon counting.

First, note that as the classical FI information is equal to
\begin{equation}
F_\Gamma=\sum_i p(i)\left(\frac{\partial_\var p(i)}{p(i)}\right)^2,
\end{equation}
two measurement results $i,j$ may be merged without loss if $(\partial_\Gamma p(i)/p(i))^2=(\partial_\Gamma p(j)/p(j))^2$.
Second, for a two-outcome measurement $p(0)=1-\epsilon \var$, $p(1)=\epsilon\var$, the term $p(0)(\partial_\var p(0)/p(0))^2\approx \epsilon^2$ is negligible compared to $p(1)(\partial_\var p(1)/p(1))^2\approx \epsilon/\varphi$ for small $\epsilon\varphi$.
Having shown this, we are ready to analyze photon losses.

First, consider the Fock input state for simplicity:
\begin{alignat}{2}
      &p(n)&&=1-tn\Gamma(1+n_E)-t(n+1)\Gamma n_E\\
         &p(n+1)&&=t(n+1)\Gamma n_E\\
          &p(n-1)&&=tn\Gamma(1+n_E).
\end{alignat}
Then indeed the outcomes $p(n-1)$, $p(n+1)$ contain the majority of information, and they may be merged without loss, as
\begin{equation}
\left(\frac{\partial_\Gamma p(n-1)}{p(n-1)}\right)^2=\left(\frac{\partial_\Gamma p(n+1)}{p(n+1)}\right)^2=\frac{1}{\Gamma},
\end{equation}
so that the FI becomes
\begin{equation}
F_\Gamma\approx \frac{p(n-1)+p(n+1)}{\Gamma}=\frac{n(1+2n_E)+n_E}{\Gamma}.
\end{equation}

Next, note that for the squeezed vacuum input state, for any odd $n$ the ratio $(\partial_\Gamma p(n)/p(n))^2$ will still be equal to $1/\Gamma$.
The total probability of getting an odd number of photons will be the sum of the probability of losing one photon and taking one photon from the environment, which are given as:
\begin{equation}
\begin{split}
p_{\t{losing}}&=\sum_{n}p_{\t{in}}(n)n\Gamma t(1+n_E)\\
p_{\t{taking}}&=\sum_{n}p_{\t{in}}(n)(n+1)\Gamma t n_E
\end{split}
\end{equation}
so finally $p_{\t{odd}}=\Gamma t [N(2n_E+1)+n_E]$ and $F_\Gamma\approx t [N(2n_E+1)+n_E]/\Gamma$.

One can note that for very large $k$ the probability of obtaining $p(2k+1)$ cannot be approximating at first order as $p(2k+1)\approx t(2k+1)\Gamma (1+n_E)p_{\t{in}}(2k)+t(2k+2)\Gamma n_Ep_{\t{in}}(2k+2)$, since the condition 
$(k+1)\Gamma t(n_E+1)\ll 1$ will not be satisfied.
Nonetheless, its impact will be negligible, since $p_{\t{in}}(2k)$ decreases exponentially with increasing $k$.

\section{Temperature estimation}
\label{app:temperature}
Here we discuss the problem of temperature estimation, comparing the results to the bounds existing in the literature and investigating details of the optimal strategy based on the usage of Fock states.

First, we recall that the bound from~\cite{wang2020quantum,shi2023ultimate} is derived only for passive strategies, i.e. ones involving preparation of the input state, free evolution for a time $t$ without any additional actions, and a final measurement.
The master equation \eqref{eq:model} with $H=0$, after interrogation over time $t$, is formally equivalent to mixing mode $a$ with a thermal mode $b$ via a beamsplitter with transitivity $\kappa=e^{-t\Gamma}$, i.e.:
\begin{equation}
\begin{split}
    \hat a'&=\sqrt{\kappa}\hat a+\sqrt{1-\kappa}\hat b,\\
    \hat b'&=-\sqrt{1-\kappa}\hat a+\sqrt{\kappa}\hat b.
    \end{split}
\end{equation}
For such a formulated problem, the QFI of the output state may be bounded using the method based on channel purification, which gives~\cite{wang2020quantum,shi2023ultimate}\footnote{Note that in \cite{wang2020quantum} some typos appear in the finite formula in App. A. Correct results may be obtained by substituting Eqs. (4), (5) into Eq. (A4).
In~\cite{shi2023ultimate} the formula concerns another parameter $n_B=(1-\kappa)n_E$, so 
$F_{n_E}=F_{n_B}(1-\kappa)^2$}:
\begin{equation}
\label{eq:shi}
I_{n_E}\leq    \frac{1}{n_E(n_E+\frac{1}{1-\kappa})}+\frac{\kappa N(2n_E+1)(1-\kappa)}{n_E(n_E+1)(n_E(1-\kappa)+1)^2}.
\end{equation}
By direct calculation, one can see that in the limit of short times $(1+n_E)t\Gamma\ll 1$ it converges to \eqref{eq:temp}. 


\textbf{Technical details of the computation of the FI for Fock states.}
Starting with a twin-Fock state $\ket{m,n}$, the probability of obtaining the twin-Fock state $\ket{m',n'}$ after going through beam splitter with transitivity $\kappa$ is given as \cite{kim2002entanglement}:
\begin{multline}
    p(n',m'|n,m)=\frac{n'!m'!}{n!m!}
    \bigg[\sum_{i=0}^n\sum_{j=0}^m\binom{n}{i}\binom{m}{j}\\
    (-1)^{j}\sqrt{\kappa}^{n+m-i-j}\sqrt{1-\kappa}^{i+j}\delta_{n-i+j,n'}\delta_{m-j+i,m'}\bigg]^2,
\end{multline}
where two last deltas imply, that non-zero value may be obtained only for $n'+m'=n+m$. For shorter notation, we define therefore:
\begin{equation}
   p(n'|n,m)=\begin{cases}
   p(n',n+m-n'|n,m),\quad &\t{if}\, n'-n\leq m\\
   0,&\t{otherwise}.
   \end{cases}
\end{equation}
As the thermal state is a mixture (not superposition) of Fock states $|m \rangle \langle m |$ with probabilities $p_{n_E}(m)=\frac{n_E^m}{(n_E+1)^{m+1}}$, by tracing out the environment we have:
\begin{equation}
    p(n'|n)=\sum_{m=0}^{\infty}  p_{n_E}(m)p(n'|n,m).
\end{equation}
The classical FI is equal to $\sum_{n'=0}^\infty (\partial_{n_E}p(n'|n))^2/p(n'|n)$.
Note that the dependence on $n_E$ lies only in $p_{n_E}(m)$, and therefore:
\begin{equation}
\label{eq:clasfish}
    F_{n_E}=\sum_{n'=0}^\infty\frac{(\sum_{m=0}^{\infty}  [\partial_{n_E}p_{n_E}(m)]p(n'|n,m))^2}{\sum_{m=0}^{\infty}  p_{n_E}(m)p(n'|n,m)}.
\end{equation}
For low temperatures, the sums converge rather quickly, so they may be effectively computed by introducing a numerical cutoff.





\section{Random displacement with Gaussian distribution vs. temperature estimation}
\label{app:axion}

Here we discuss the relation between random displacement with Gaussian distribution and temperature estimation.
Let us come back to the problem of displacement estimation, but assume that neither intensity more direction of displacement $\alpha$ is fixed, but $\alpha$ is a random complex Gaussian variable with zero mean $p(\alpha)=\frac{1}{\pi \sigma^2}\exp({-|\alpha|^2/\sigma^2})$ and the aim is to estimate the variance of these distribution $\sigma^2$.
For a fixed time (i.e. time dependence was not taken into account), such a problem has been discussed in~\cite{shi2023ultimate} in the context of dark matter sensing.
Let us recall the crucial observations.

Consider the quantum channel consisting displacement $\beta$ and mixing with thermal mode $b_{n_E}$ (where $n_E$ is mean number of thermal photons):
\begin{equation}
a'= \sqrt{\kappa}a+\beta+\sqrt{1-\kappa}b_{n_E}.
\end{equation}
If $\beta$ is a random complex variable with Gaussian distribution $p(\beta)=\frac{1}{\pi \sigma^2}\exp(-|\beta|^2/\sigma^2)$, this channel is exactly equivalent to mixing with thermal mode with the appropriate number of thermal photons:
\begin{equation}
a'= \sqrt{\kappa}a+\sqrt{1-\kappa}b_{n_E+n_\beta},\quad \t{where}\quad n_\beta=\frac{\sigma^2}{1-\kappa}
\end{equation}
Now we are ready to consider the time dependence.
The master equation \eqref{eq:model} with the displacement Hamiltonian \eqref{eq:dispH}, for any fixed $\alpha$, leads to the mode transformation:
\begin{equation}
a'= e^{-t\Gamma/2}a+2\alpha(1-e^{-t\Gamma/2})/(\Gamma/2)+\sqrt{1-e^{-t\Gamma}}b_{n_E}.
\end{equation}
If $\alpha$ is a random complex variable with distribution $p(\alpha)=\frac{1}{\pi \sigma^2}\exp({-|\alpha|^2/\sigma^2})$, this is equivalent to:
\begin{equation}
a'= e^{-t\Gamma/2}a+\sqrt{1-e^{-t\Gamma}}b_{n_E+n_\alpha(t)}
\end{equation}
where:
\begin{equation}
n_\alpha(t)=\frac{\sigma^2 \left[2\alpha(1-e^{-t\Gamma/2})/(\Gamma/2)\right]^2}{1-e^{-t\Gamma}}.
\end{equation}
We see that for any fixed time, we may map the channel obtained in this way to a channel mixing the original mode with a thermal bath.
However, the effective number of thermal photons $n_E+n_A(t)$ is a function of time, therefore conclusions about the optimal time-dependent strategy will be substantially different.
Specifically, for short times $n_\alpha(t)$ grows quadratically with $t$, so, unlike in temperature estimation, the fast-prepare-and-measure protocol will be far from optimal.

We should stress that in the above calculation, perfect coherence in time of the variable $\alpha$ is assumed (i.e. once it is randomly drawn from the Gaussian distribution, it remains fixed for any time).
However,iIn the context of dark matter sensing, also changes of $\alpha$ in time should be included.
See~\cite{shi2024quantum} for a realistic approach to time-dependence in this problem.

\bibliography{biblio}

\end{document}